# Neuronal arithmetic operators based on Ovonic threshold switches (OTS) for biologically inspired analog computing

Jingyeong Hwang[1,2], Jaesang Lee[1], Jiin Bang[1,3], Younghyun Lee[1], Unhyeon Kang[1,2], Seungmin Oh[1,4], Kyungmin Lee[1,5], Jaehyun Park[1,6], Seongsik Park[1], Hyun Jae Jang[1], Sangbum Kim[2], Min Hyuk Park[2], Suyoun Lee[1, 3]*

[1]*Center for Semiconductor Technology, Korea Institute of Science and Technology, Seoul 02792, Korea*

[2]*Department of Materials Science and Engineering, Seoul National University, Seoul 08826, Korea*

[3]*Nanoscience and Technology, KIST School, University of Science and Technology, Seoul 02792, Korea*

[4]*Department of Physics and Astronomy, Seoul National University, Seoul 08826, Korea*

[5]*Department of Electrical Engineering, Korea University, Seoul 02841, Korea*

[6]*Department of Materials Science and Engineering, Korea University, Seoul 02841, Korea*

*Email: S. Lee (slee_eels@kist.re.kr)

**Biological neurons perform arithmetic computations – including additive integration and divisive gain modulation – through synaptic conductance changes and shunting inhibition, enabling context-dependent information processing that far exceeds simple threshold-and-fire models. Replicating these capabilities in compact hardware remains a fundamental challenge for neuromorphic engineering. Here, we demonstrate artificial neuron circuits based on Ovonic threshold switches (OTS) that physically implement three arithmetic operations: SUM, PARALLEL, and DIVISION. The SUM and PARALLEL neurons exploit MOSFET-controlled dendritic conductances, producing output firing rates that collapse onto invariant curves as a function of combined inputs — satisfying the canonical criteria for neuronal addition. The DIVISION neuron**

**leverages a JFET-based shunting pathway, inspired by GABA$_A$-mediated inhibition in the cortex, to achieve divisive gain modulation well described by a Hill-type function ($R^2 \approx 0.95$, Hill exponent n ≈ 1.3), consistent with nonlinear normalization observed in visual and olfactory circuits. Applying the DIVISION neuron to pixel-wise image normalization under non-uniform illumination recovers obscured visual content, mirroring contrast normalization in the visual cortex. Compared to CMOS-based division implementations, the proposed approach offers improvements in energy efficiency and scalability exceeding an order of magnitude, establishing a viable path toward compact, brain-inspired analog computing.**

## Introduction

The rapid growth of artificial intelligence (AI) has exposed fundamental limitations of conventional digital computing architectures, particularly the energy cost and data-movement overhead associated with the von Neumann bottleneck.[1,2] Modern AI workloads rely on a myriad of digital operations to mimic arithmetic operations in neural networks, motivating increasing interest in analog devices that can perform computation directly in the physical domain, thereby enabling in-memory and near-sensor computing.[2-4] Such approaches exploit device-level physics, including Ohmic conduction, current summation, and nonlinear conductance modulation, to carry out computation with potentially lower power consumption and latency than digital implementations.[5]

On the other hand, biological neural systems offer key insights for the design of energy-efficient analog computing devices. Neurons process information not only through integration and thresholding, but also by dynamically transforming synaptic inputs.[6,7] For example, in the presence of multiple synaptic inputs, some neurons show the input-output (I-O) relationship systematically varying with the combination of inputs, resulting in a parallel shift or a change in the slope of the I-O curve as shown in Fig. 1a.[8] As articulated by *Silver* (2010), such systematic changes in their I-O curves underpin arithmetic operations in single neurons, particularly in the cerebral cortex.[7,9] In this context, a parallel shift of the I–O curve corresponds to additive or subtractive computation, whereas a change in slope reflects multiplicative or divisive modulation.[9-12]

These transformations arise from diverse synaptic and neuromodulatory mechanisms, including conductance changes induced by neuromodulators and fast inhibitory processes such as *shunting inhibition* mediated by $GABA_A$ ($\gamma$-aminobutyric acid type A) receptors.[10,13-17] In particular, neuromodulator-driven conductance changes - mediated by substances such as acetylcholine and serotonin[18] - primarily shift the I-O curve, while fast synaptic processes such as shunting inhibition via $GABA_A$ receptors primarily modulate its slope.[10,13-17] Together, these mechanisms enable flexible and context-dependent transformations of synaptic inputs in biological neurons.

These biological principles have inspired the development of neuromorphic arithmetic operators for compact and energy-efficient hardware computation. Previous device-level studies, including $NbO_x$-based memristor neurons[19] and graphene oxide-coupled electrolyte-gated neuron transistors[20], have reported input-dependent modulation of I-O curves. Nevertheless, in those works, it was not demonstrated that the output of the operation is a single-valued function of multiple inputs, irrespective of the combination of the inputs, essential requirement for the device to act as an operator. In this work, based on those operation principles of biological neurons, we introduce devices that can conduct analog arithmetic SUM, PARALLEL, and DIVISION operations. Each device consists of a threshold switch device, which outputs an analog signal in the form of the firing rate in the rate-coding scheme, and a few JFETs (junction field-effect transistors) as a voltage-controlled resistor, which enables us to modulate input signals in the analog domain. It is shown that these devices can mimic the behaviors of biological neural operators shown in Fig. 1a, and the firing rate is exactly mapped to the output of the arithmetic operations, irrespective of the combination of the inputs. Furthermore, we demonstrate the functional relevance of our approach by applying the DIVISION operator to an image-division task, a key computation in the visual cortex, thereby enabling normalization under non-uniform illumination.

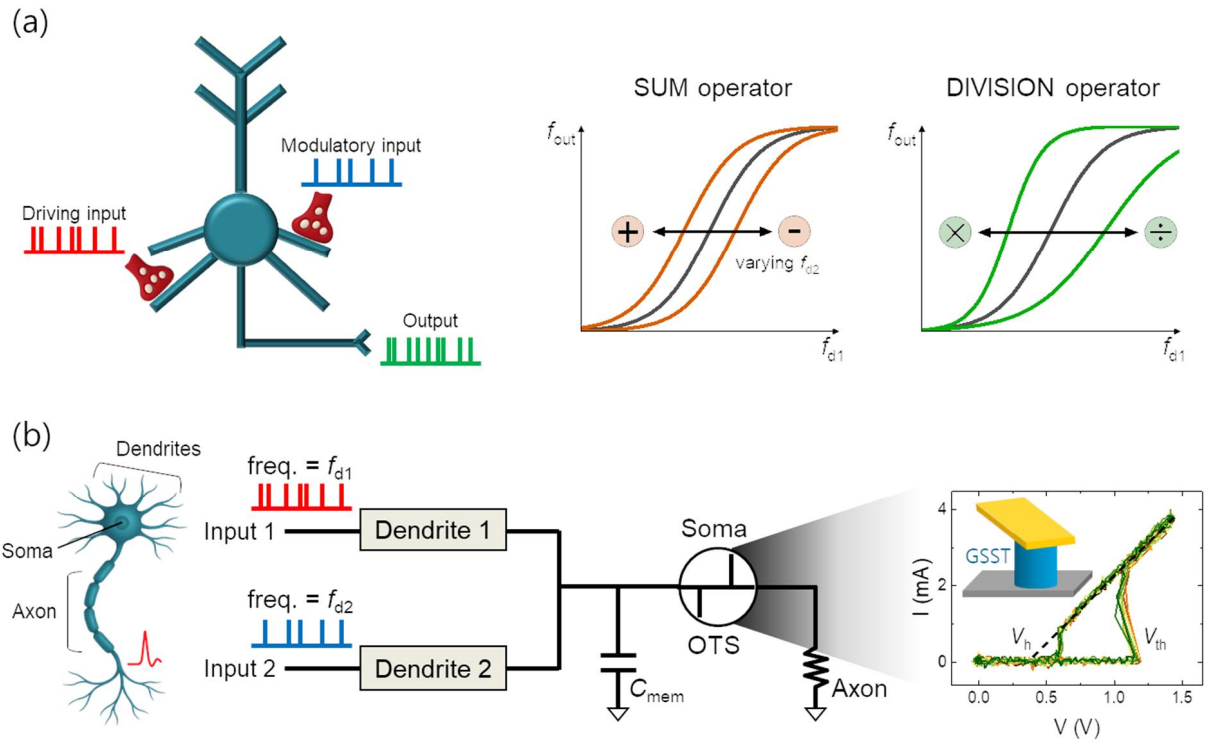

**Figure 1 Neuromorphic arithmetic operator based on Ovonic Threshold Switch (OTS).** (a) Schematic illustration of neuronal arithmetic under a rate-coding paradigm. A neuron receives a driving input (red) and a modulatory input (blue) that primarily determines spiking activity and regulates neuronal responsiveness without directly triggering spikes, respectively. Depending on the modulation mechanism, the input-output (I-O) firing-rate curve can exhibit an additive shift along the input axis (left, additive modulation) or a change in slope corresponding to gain control (right, divisive modulation). (b) Circuit-level implementation of an OTS-based artificial neuron. Two independent inputs are applied through dendritic branches to an OTS device acting as the spiking element, followed by an axonal load resistor. The circuit abstracts biological dendritic integration and somatic thresholding.

# Results

## SUM and PARALLEL operators based on dendrite conductance modulation

Fig. 2a shows the circuit that implements the SUM operation. The basic 1 OTS + 1 C + 2 R structure of a neuron was introduced in our previous work,[21] and here the original dendritic resistor is replaced by two parallel-connected n-type MOSFETs. The OTS device in this work is fabricated with Pt/Ge$_2$Se$_2$Sb$_1$Te$_{1.2}$/TiN stack in a pore-type structure (see Methods section and Supplementary Information Figure S1 for details). Each FET receives an independent gate voltage input ($V_{in1}$, $V_{in2}$), which controls its channel resistance, and the resulting composite resistance ($R_{comp}$) determines the RC charging delay of the membrane capacitor ($C_{mem}$). Consequently, the output firing rate ($f_{out}$) is given by:

$$f_{out} \propto \frac{1}{C_{mem}R_{comp}} = \frac{1}{C}\left(\frac{1}{R_{d1}} + \frac{1}{R_{d2}}\right) = \frac{1}{C}\{G_{d1}(V_{in1}) + G_{d2}(V_{in2})\} \qquad \ldots(1)$$

Here, $R_i$ ($i$=d1 and d2) denotes the resistance of the $i$-th input channel. Fig. 2b presents representative temporal waveforms of the input voltages and the neuron output. As $V_{in1}$ and $V_{in2}$ are increased stepwise, the output spike frequency increases accordingly. Notably, both the firing rate and the spike offset voltage increase monotonically with input voltage, consistent with enhanced charging dynamics due to increased input conductance. Fig. 2c shows a linear relationship between the output firing rate ($f_{out}$) and the summed input voltage ($V_{in1}+V_{in2}$), demonstrating SUM arithmetic behavior in the circuit.

To further validate the correspondence of the device to the biological SUM operation in the rate-coding scheme, we have investigated its I-O transformation using Poisson-distributed pulse trains. Fig. 2d shows the output firing rate as a function of the primary input frequency ($f_{in1}$), while systematically varying the secondary input frequency ($f_{in2}$) across three conditions:

0, 0.2, and 0.4 MHz. As $f_{in2}$ increases, the I-O curve is parallel-shifted to the left (lower $f_{in1}$ region) without any change in slope around 1.4. As mentioned above, this horizontal parallel shift of the I-O curve is a hallmark of additive neuronal integration. Crucially, when replotted as a function of the summed input frequency ($f_{in1}+f_{in2}$), all data collapse onto a single straight line, indicating that the neuron response is primarily determined by the total input rate. Together, these results satisfy the canonical criteria for neuronal addition, confirming that our OTS-based SUM neuron performs genuine additive arithmetic behavior.

The PARALLEL operator shown in Fig. 2f differs from the SUM configuration in that the two n-channel FETs are connected in series rather than in parallel. This configuration results in an inverse-additive relation in the effective input pathway. In this case, $f_{out}^{PARA}$ is given by the following expression.

$$\frac{1}{f_{out}^{PARA}} \propto CR_{comp} = C(R_{d1} + R_{d2}) = C\left(\frac{1}{G_{d1}} + \frac{1}{G_{d2}}\right) \quad \ldots(2)$$

Fig. 2g presents the temporal output waveforms measured under stepwise increases of $V_{in1}$ and $V_{in2}$. Following Eq. (2), $1/f_{out}$ is plotted as a function of ($1/V_{in1} + 1/V_{in2}$) in Fig. 2h, revealing a linear dependence characteristic of the PARALLEL operation. When used in conjunction with the SUM operator, this PARALLEL operator can be employed to extract multiplicative relationships between the inputs.

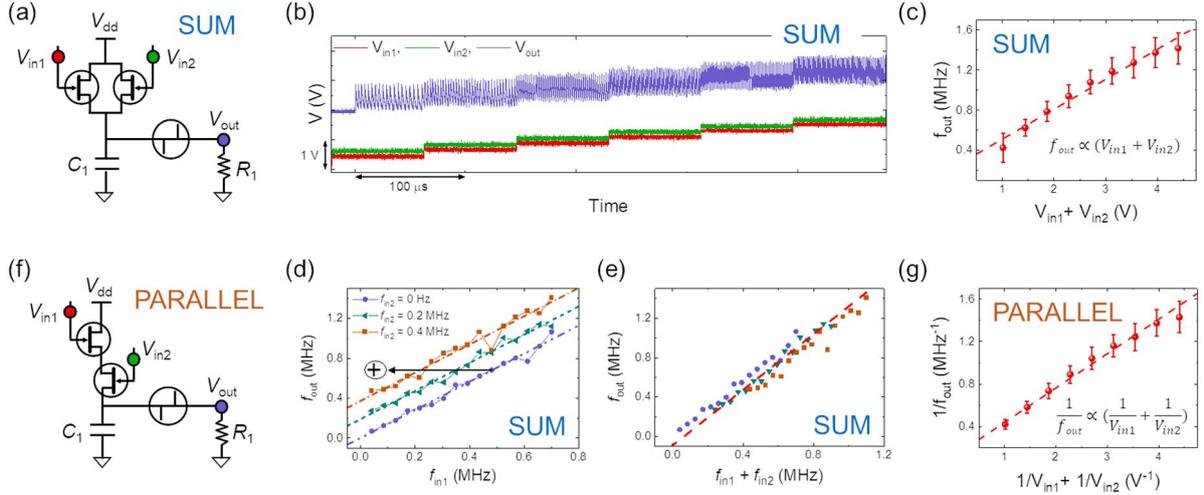

**Figure 2 SUM and PARALLEL arithmetic neural operators.** (a) Circuit diagram of the neural SUM operator based on OTS, (b) Waveforms of $V_{in1}$, $V_{in2}$ and $V_{out}$ for SUM operator. (c) The firing rate of the output spikes ($f_{out}$) as a function of $V_{in1}+V_{in2}$ for the SUM operator. (d) $f_{out}$ as a function of $f_{in1}$ with varying $f_{in2}$ for the SUM operator, where $f_{in,i}$ ($i$=1,2) is the firing rate of the Poisson pulse train inputs. (e) $f_{out}$ as a function of $f_{in1}+f_{in2}$, indicating the translation-invariant property of the I-O curve. (f) Circuit diagram of the PARALLEL neural operator. (g) Reciprocal of $f_{out}$ as a function of $1/V_{in1}+1/V_{in2}$ for the PARALLEL operator. The error bar in (c) and (g) represents a standard deviation obtained from 10 repetitions. A dashed line is a linear fit to the data.

## DIVISION operator based on shunting inhibition

In addition to basic SUM and PARALLEL operations, MULTIPLICATION and DIVISION are essential operations for efficient information processing in the brain. For example, in cortical neurons, the divisive gain modulation relies on the shunting inhibition mediated by $GABA_A$ receptor, which activates shunting conductance near the soma.[11,14] This reduces neuronal gain by dynamically scaling excitatory responses, thereby modulating the slope of the I-O relationship (Fig. 1a). Such an operation principle can be implemented by combining a basic neuron with a variable shunt resistor controlled by a modulatory signal. Here, we implement this concept using an OTS-based neuron in which a JFET serves as a voltage-controlled resistor. Fig. 3a shows the circuit of the proposed OTS-based DIVISION operator, where a shunt branch consists of an n-channel JFET (2N4393, InterFET), two resistors ($R_2$, $R_3$), and a capacitor ($C_2$), connected in parallel to the membrane node. As $V_G$ increases, the JFET conductance increases, diverting current away from the soma-equivalent node and reducing the effective voltage across the OTS device. This suppresses spike initiation and lowers the output firing rate, consistent with divisive gain control. It is noteworthy that the structure of the shunting path closely resembles that of the basic neuron, with the OTS element replaced by a JFET in the shunt branch. This similarity suggests that a range of functionalities can be realized by combining units that share a common structural framework but differ in their switching elements - an approach reminiscent of the diversity of biological neurons, which exhibit varied functions despite a largely conserved architecture.

To examine the detailed behavior of the DIVISION neuron, we applied driving and modulatory inputs in the form of Poisson-distributed pulse trains. The average firing rates of the driving and modulating pulse trains ($f_{dr}$ and $f_{mod}$, respectively) were varied, while the pulse amplitude and width were fixed at 6 V and 500 ns. Fig. 3b shows that activation of the shunting path suppresses the output spikes. Specifically, as the modulatory gate voltage $V_G$ gradually increases and exceeds the pinch-off threshold $V_p$, spike generation is progressively attenuated, indicating effective shunting-induced suppression of firing activity. Fig. 3c shows representative waveforms under a fixed driving input condition ($f_{dr}$=0.4 MHz) for two different modulatory inputs: $f_{mod}$ = 0.32 MHz and 0.725 MHz, respectively. It is clearly observed that the firing rate of the output spikes is lower in the high modulation case ($f_{mod}$ = 0.725 MHz) than in the low modulation case ($f_{mod}$ = 0.32 MHz). Using fast Fourier transform (FFT) of the $V_{out}$ waveform, we extract $f_{out}$ as a function of $f_{dr}$ and $f_{mod}$, as shown in Fig. 3d. It is observed that increasing $f_{mod}$ results in a progressive reduction in slope, realizing the divisive gain modulation of a biological cell shown in Fig. 1a. Most importantly, as shown in Fig. 3e, when $f_{out}$ is plotted as a function of $f_{dr}/f_{mod}$, the response curves collapse onto a single unified curve with minimal scatter. We found that the conductance range of the shunting path significantly influences the response curves, and that optimizing this range is critical for achieving scale-invariant behavior of the DIVISION operator (see Section S2, "Optimization of the Operation Window of the DIVISION Operator," in the Supplementary Information). Under optimized conditions, the observed scale invariance indicates that the proposed circuit performs a true DIVISION operation within a rate-coding framework.

To model the neuron's divisive behavior more quantitatively, we fitted the observed dependence of $f_{out}$ on $f_{dr}/f_{mod}$ using a Hill-type function, $H(x)$, which captures sigmoidal input-output relations in receptor-ligand binding kinetics.[22-24]

$$f_{out}(x) = f_{max}H(x) = f_{max}\frac{x^n}{x^n + x_{50}^n} \quad \text{(where } x = \frac{f_{dr}}{f_{mod}}) \quad \ldots(3)$$

Here, $f_{max}$, $x_{50}$, $f_{out}(x_{50})$, and $n$ are the maximum output firing rate, the input ratio that yields a half-maximal response, $0.5*f_{max}$, and the Hill exponent that determines the steepness of the response, respectively. The fitting curve is shown as a dashed line in Fig. 3e, releasing $R^2$ of ~ 0.95. It indicates that the model captures the measured nonlinear dynamics over a range of input ratios. Notably, the extracted Hill exponent in our OTS-based DIVISION neuron is estimated at $n \approx 1.3$. The fact that $n > 1$ indicates a nonlinear divisive normalization effect, which is qualitatively consistent with previous observations in biological neurons. In the visual cortex, contrast response functions are well described by hyperbolic ratio functions with exponents empirically greater than unity ($n > 1$) and, in olfactory projection neurons, a similar nonlinear gain modulation was reported with $n \approx 1.5$.[25-28] Taken together, these results support the interpretation that the proposed hardware captures key features of nonlinear divisive normalization observed in sensory circuits.[29]

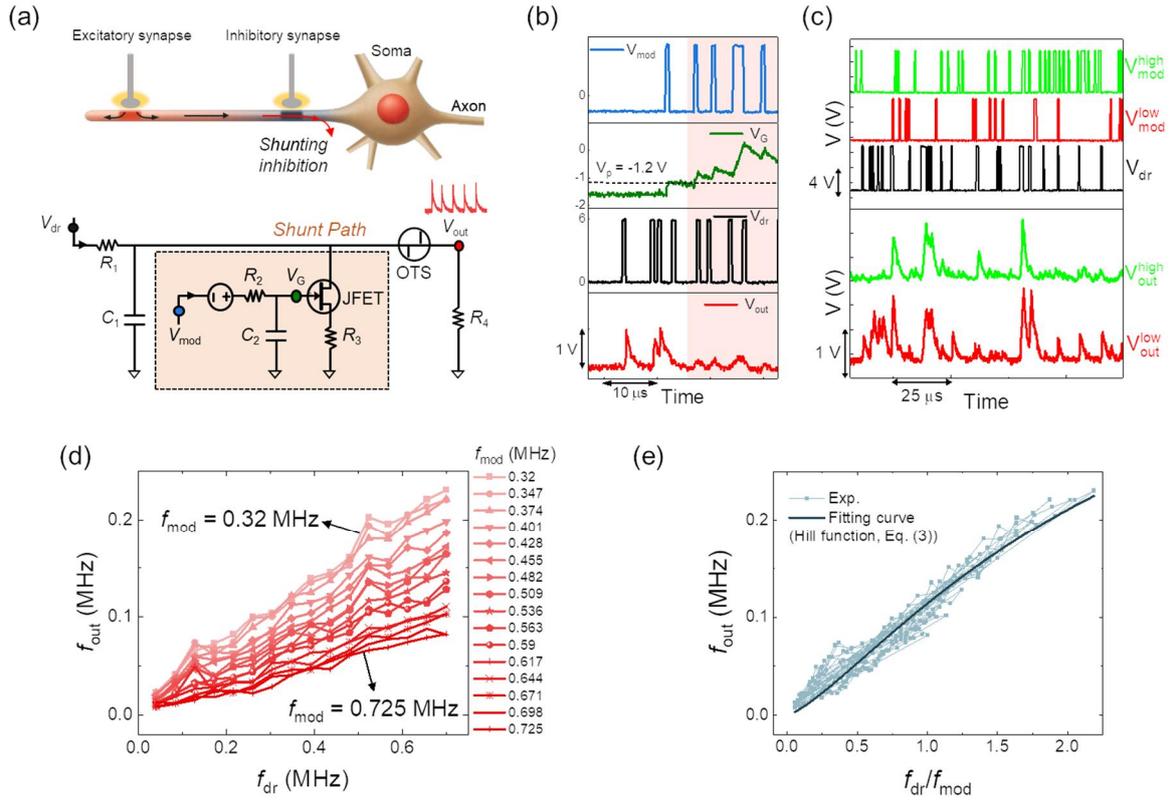

**Figure 3 Rate-coded implementation of DIVISION arithmetic operator based on shunting inhibition.** (a) Circuit diagram of the DIVISION neuron featuring a voltage-controlled shunt path. The shunting pathway includes a n-channel JFET (2N4393, InterFET) acting as a voltage-controlled resistor, modulated by gate voltage $V_G$, which is determined by $f_{mod}$ through a low-pass filtered input. Component values: $R_1$ = 4.7 kΩ, $R_2$ = 1 kΩ, $R_3$ = 500 Ω, $R_L$ = 5.1 kΩ, $C_1$ = 82 pF, and $C_2$ = 5 nF. The driving input $f_{dr}$ and modulatory input $f_{mod}$ are applied as a Poisson-distributed spike trains. Input spike trains used a 6 V amplitude and 500 ns width. (b) Temporal waveforms showing spike suppression via shunting inhibition. As $V_G$ increases and exceeds the pinch-off threshold $V_p$, the output spike activity becomes progressively suppressed. (c) Output spike trains recorded under identical driving input (black, $f_{dr}$= 0.4 MHz) with two levels of modulatory input (red and green): 0.32 MHz and 0.725 MHz, respectively. Stronger modulation leads to more pronounced suppression of spiking activity, confirming input-dependent gain control. (d) Output firing rate ($f_{out}$) as a function of $f_{dr}$ for different $f_{mod}$, showing slope reduction with stronger inhibitory input. (e) $f_{out}$ of a DIVISION operator as a function of $f_{dr}/f_{mod}$, demonstrating the scaling-invariant property of the operator.

**Real-time Image Contrast Enhancement using the DIVISION operator**

A major challenge in vision systems is maintaining consistent image contrast under spatially non-uniform illumination. The human visual system addresses this problem through divisive normalization, in which neuronal sensitivity is adaptively scaled according to the surrounding context.[25,27,29,30] To examine the practical utility of the proposed DIVISION neuron, we applied its measured response to an image-normalization task. Specifically, we performed pixel-wise division of an original image by a background image acquired under the same non-uniform illumination, which is analogous to divisive normalization and the Naka-Rushton model[26].

For this task, we prepared an 8-bit gray-scale original image containing text under non-uniform illumination, as shown in Fig. 4a. We obtained the background image directly from the original image by using a local background-estimation step based on morphological closing[31,32], as shown in Fig. 4b. This single-image approach provides a practical normalization scenario, in which the denominator is derived from the spatial context of the same image. Notably, the gap-filling logic of morphological closing finds a functional analogue in cortical visual processing: deep-layer V1 neurons reconstruct missing signals at the retinal blind spot via lateral activity propagation[33,34], mirroring the dilation step, while LOC-to-V1/V2 feedback subsequently refines local boundary details from a globally integrated shape representation[35], paralleling the erosion step. A 256×256 lookup table was then generated from the measured DIVISION-neuron response by interpolation, providing $f_{out}$ as a function of $f_{dr}$ and $f_{mod}$ (Fig. 4c). For each pixel, the grayscale values of the original and background images were linearly mapped to $f_{dr}$ and $f_{mod}$, respectively, and the corresponding $f_{out}$ value was retrieved from the lookup table. The corresponding division ratio ($f_{dr}/f_{mod}$) was then obtained through the inverse Hill relation, $f_{dr}/f_{mod}=H^{-1}(f_{out})$, based on Eq. (3). This ratio was finally remapped to an 8-bit grayscale value in the range of 0~255. As shown in Fig. 4d, the processing evens out illumination and reveals text in shadowed areas without degrading the overall image structure. These results provide a proof-of-concept demonstration that the measured response of the OTS-based DIVISION neuron can be translated into an image-normalization function relevant to bioinspired vision processing.

To further assess the practical efficiency of the proposed DIVISION neuron, we estimated the electrical energy consumed during a single rate-coded division operation. Here, the consumed energy ($E_{pixel}$) was estimated by time-integrating the instantaneous electrical power in both input paths ( $E_{pixel} = \sum_{i=dr,mod} \int_0^T V_i(t)I_i(t)dt$ , see Fig. S3b in Supplementary Information), with one pixel-wise division being defined as the application of a pair of Poisson-distributed driving and modulatory pulse trains over a fixed time window ($T$). To represent 256 gray levels within the rate-coding scheme, $T$ is set to 6.4 ms, corresponding to 256 times the period of the slowest driving pulse train in this experiment (= 25 μs). Since $E_{pixel}$ depends on the input firing rates, four representative input combinations spanning the minimum and maximum values of $f_{dr}$ and $f_{mod}$ were measured, and their average was used as the representative $E_{pixel}$ for the division operation. Under the condition of $V_{dr} = V_{mod} = 2$ V and $R_1 = 1$ kΩ, the average energy per rate-coded division operation was measured to be ~37.4 μJ (see Fig. S3c in Supplementary Information). We further compared the proposed DIVISION neuron with previously reported CMOS-based division implementations from the viewpoint of energy consumption and circuit complexity. Previous studies have suggested that the energy consumption of OTS-based neurons approximately follows a device-scaling relation proportional to ~ $d^{1.6}$, where $d$ denotes the characteristic dimension of the OTS device.[36] When this relation is used only in an indicative sense, the present micrometer-scale result corresponds to an estimated energy on the order of ~pJ per division operation at a nominal device dimension of 16 nm.[24] This estimate is not intended as a quantitative prediction, but rather as an order-of-magnitude reference to contextualize the measured result.

Reported CMOS-based division operators span a broad range of architectural strategies and implementation costs. For instance, a restoring sequential 8-bit divider implemented in 45-nm CMOS consumes 100 ~ 1000 pJ per division and requires a few hundred transistors. An asynchronous divider based on Algorithm H, implemented in 90-nm CMOS, reports an energy consumption of approximately 180 pJ per operation, with a circuit complexity exceeding 10,000 transistors.[25] More recently, a signed-array 8-bit divider implemented as a CMOS ASIC has achieved energy consumption of a few pJ per division, while still relying on several thousand transistors.[26]

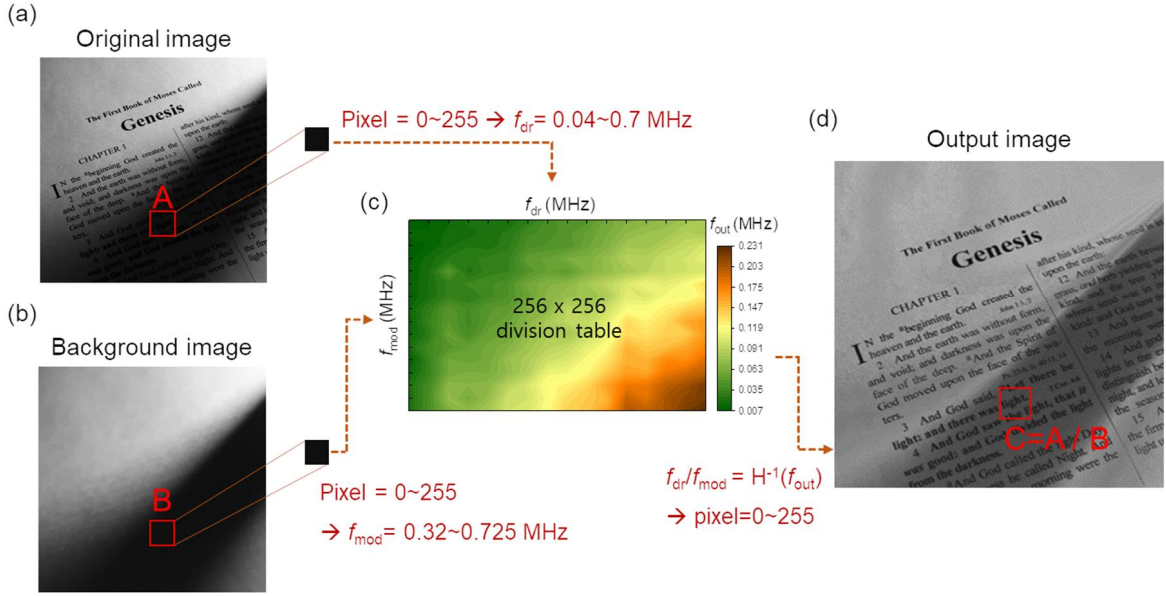

**Figure 4 Pixel-wise image division using the OTS-based DIVISION neuron.** (a, b) Original image containing textual content under non-uniform illumination and background image obtained directly from the original image using morphological closing. (c) Experimentally measured output firing rates of the DIVISION neuron for combinations of driving ($f_{dr}$, mapped from pixel values 0-255 to frequencies of 0.04-0.7 MHz) and modulatory ($f_{mod}$, mapped to frequencies of 0.32-0.725 MHz) inputs. This firing rate table serves as a lookup for the analog division process, converting the neuron's output back into pixel values through the inverse Hill function. (d) Resulting image after division (C=A/B), demonstrating significantly improved visibility and illumination uniformity.

## Discussion

In this work, we presented three OTS-based neuronal arithmetic circuits that implement SUM, PARALLEL, and DIVISION operations, targeting improved scalability and energy efficiency in neuromorphic computing systems. Each circuit exhibits a characteristic collapse of input–output data onto invariant curves, confirming its genuine arithmetic functionality. In particular, the DIVISION neuron demonstrates gain-modulatory behavior, realized through a shunting inhibition mechanism inspired by neurons in the visual cortex.

Furthermore, scale-invariant operation is verified by the emergence of a unified response as a function of $f_{dr}/f_{mod}$, which can be well described by a Hill-type model. By applying the measured DIVISION-neuron response to a pixel-wise image normalization task, we demonstrate reduced illumination non-uniformity and enhanced visibility in previously obscured regions. Finally, through comparison with CMOS-based counterparts, we confirm that the proposed OTS-based neuronal arithmetic operators achieve improvements in both scalability and energy efficiency by more than an order of magnitude.

In conclusion, the development of highly scalable and energy-efficient neuronal arithmetic operators presented in this work paves the way for a new class of computing architectures that bridge the gap between biological computation and conventional digital processing. By leveraging intrinsic device dynamics to realize arithmetic functionalities, the proposed approach enables compact, low-power implementations that can be readily extended to large-scale systems. This advancement holds significant promise for next-generation edge intelligence, real-time signal processing, and adaptive learning platforms, where both computational density and energy constraints are critical. Moreover, the compatibility of the proposed operators with emerging memory and neuromorphic devices suggests a viable pathway toward fully integrated, brain-inspired hardware systems, potentially transforming the landscape of efficient computing beyond the limits of traditional CMOS technologies.

# Methods

**Fabrication and electrical characterization**

The OTS device used in this study consists of an amorphous chalcogenide active switching layer sandwiched between metallic top and bottom electrodes. Figure S1 presents a cross-sectional transmission electron microscope (TEM) image of the device. A pore-type structure was employed to confine current flow through the active material. Specifically, the bottom electrode (BE), composed of Ti (5 nm)/Pt (100 nm), was patterned on a $SiO_2$ substrate via a lift-off process. A 70 nm-thick insulating $SiO_2$ layer was subsequently deposited by plasma-enhanced chemical vapor deposition (PECVD). Pores with various diameters ($d$) were then formed in the $SiO_2$ layer above the BE using electron-beam lithography (for $d \leq 500$ nm) and photolithography (for $d \geq 2$ μm), followed by reactive ion etching (RIE). A 100 nm-thick $Ge_2Se_2Sb_1Te_{1.2}$ (GSST) switching layer was deposited by magnetron RF co-sputtering from GeSe and $Sb_{2.5}Te_3$ targets. The composition of the GSST film was characterized by Auger spectroscopy, as shown in Fig. S4 in the Supplementary Information. Finally, the top electrode (TE), consisting of TiN (50 nm)/Ti (5 nm)/Au (100 nm), was patterned via a lift-off process.

The electrical characteristics of the OTS devices and the neuronal arithmetic circuits were measured using a two-channel arbitrary function generator (Tektronix AFG-3102) and a four-channel oscilloscope (Tektronix DPO-5104). For the DIVISION neuron, a voltage-controlled shunt resistor was implemented using a junction field-effect transistor (JFET; 2N4393, InterFET), enabling dynamic control of the shunting conductance. Operator behavior was investigated by applying two Poisson-distributed input pulse trains with independently controlled firing rates. The output firing rate was determined by counting the number of output spikes over the duration of each paired-input stimulation window.

**Pixel division using the DIVISION neuron device**

For this task, a single 8-bit grayscale image containing text under non-uniform illumination was used as the input image. To construct the denominator image for normalization, the background/illumination component was estimated directly from the original image using morphological closing. At each pixel, the grayscale values (0–255) of the original and background images were mapped to the driving and modulation inputs, respectively, encoded as input frequencies. The corresponding pulse trains were Poisson-distributed, with firing rates linearly proportional to the pixel values and a pulse width of 500 ns. These pulse trains were generated using a two-channel arbitrary function generator (Tektronix AFG-3102) with a common trigger. The input pulse trains and the resulting output spike train were recorded using a four-channel oscilloscope (Tektronix DPO-5104). The output firing rate was determined by counting the number of output spikes for each fixed pair of input pulse trains.

The experimentally obtained firing-rate table was then fitted using the Hill function described in Eq. (3) to extract the corresponding division values ($f_{dr}/f_{mod}$). These ratios were subsequently remapped to 8-bit grayscale values (0–255) to construct a grayscale division table for image processing (Fig. 4c). For image processing, both the original image (with text) and the background image (without text but under identical illumination) were converted to grayscale using the weighted sum of RGB values (0.299R + 0.587G + 0.114B), following the standard Y′UV model. Each image was quantized into 256 grayscale levels, and the division operation was performed using a 256 × 256 lookup table obtained by interpolating the experimental 16 × 16 response matrix derived from Fig. 3d. For each pixel, the grayscale level of the original image served as the $f_{dr}$ index, while that of the background image served as the $f_{\text{mod}}$ index. The grayscale value at the intersection of these indices in the division table was assigned as the output pixel value. The image-level division process was implemented using custom Python scripts developed for this study.

## Data availability

All the data supporting the findings of this study are available within the article and its Supplementary Information.

## Code availability

The simulation results were processed using Python software. All the relevant codes are available from the corresponding author upon reasonable request.

## Acknowledgments

This work was supported by the Korea Institute of Science and Technology (KIST, Grant No. 26E0020) and by the National Research Council of Science & Technology (NST, Grant No. GTL24041-000) and National Research Foundation of Korea (NRF, Grant No. RS-2025-24533987) by the Korea government (MSIT).

## Author Contributions Statement

S.L. designed and conceived the experiments. J.G.H. and J.L. fabricated the OTS devices and performed experiments demonstrating the functions of the arithematic neural operators. J.G.H., U.K. and J.B. designed and performed the image-division task. J.G.H. and S.O. composed a Python code for performing the image-division operation by using a lookup table. Y.L., K.L., S.Kim, and M.H.P. contributed to the analysis of the spike waveforms of the neural operators.

S.P. and H.J. contributed to the design of the electrical circuitry of the neural operators. All authors discussed the data and participated in revising the manuscript.

## Competing Interests Statement

The authors declare no competing interests.

# Supplementary Information

# Neuronal arithmetic operators based on Ovonic threshold switches (OTS) for biologically inspired analog computing

Jingyeong Hwang[1,2], Jaesang Lee[1], Jiin Bang[1,3], Younghyun Lee[1], Unhyeon Kang[1,2], Seungmin Oh[1,4], Kyungmin Lee[1,5] Jaehyun Park[1,6], Seongsik Park[1], Hyun Jae Jang[1], Sangbum Kim[2], Suyoun Lee[1, 3]*

[1]*Center for Semiconductor Technology, Korea Institute of Science and Technology, Seoul 02792, Korea*

[2]*Department of Materials Science and Engineering, Seoul National University, Seoul 08826, Korea*

[3]*Nanoscience and Technology, KIST School, University of Science and Technology, Seoul 02792, Korea*

[4]*Department of Physics and Astronomy, Seoul National University, Seoul 08826, Korea*

[5]*Department of Electrical Engineering, Korea University, Seoul 02841, Korea*

[6]*Department of Materials Science and Engineering, Korea University, Seoul 02841, Korea*

**S1. Device structure and characteristics of the OTS**

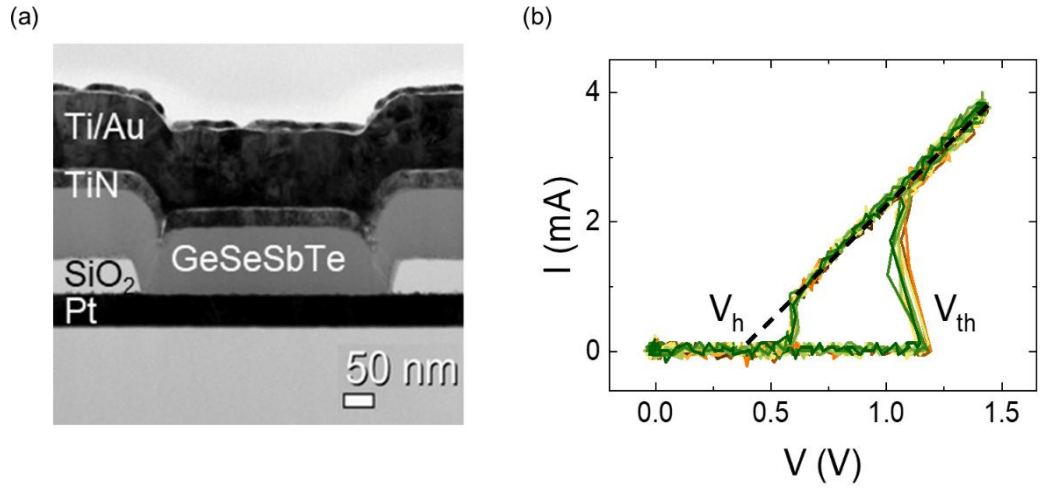

**Figure S1. Structure and switching characteristics of the OTS device.** (a) Cross-sectional TEM image of the OTS device. A pore-type structure was employed with the diameter (*d*) of the pore varying from 200 nm ~ 6 μm. A 100 nm-thick $Ge_2Se_2Sb_1Te_{1.2}$ (GSST) was used as the switching layer. (b) Characteristic switching current-voltage (*I-V*) curve of the OTS, measured with a 200 Ω load resistor connected in series. It shows ten repetitions of measurements, demonstrating good reproducibility of our OTS device.

## S2. Optimization of the operation window of the DIVISION operator

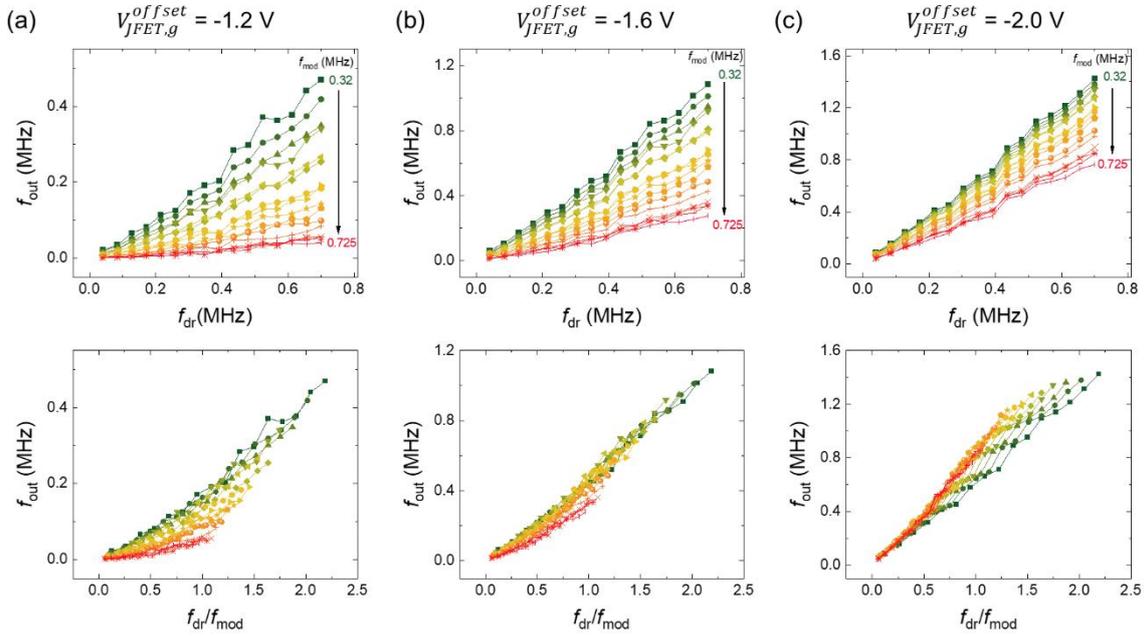

**Figure S2. Effect of the shunting conductance range of the DIVISION operator on the response curves,** (top panels) $f_{out}$ of the DIVISION operator as a function of $f_{dr}$ with varying $f_{mod}$ when the offset voltage applied to the gate node of the JFET ($V_{JFET,g}^{offset}$) is -1.2 V (a), -1.6 (b), and -2.0 V (c), respectively. (bottom panels) $f_{out}$ as a function of $f_{dr} / f_{mod}$ for each case.

We have investigated the effect of the conductance range of the shunting path ($G_{shunt}$) on the response curves of the DIVISION operator. The range of $G_{shunt}$ can be tuned by applying an offset voltage to the JFET gate ($V_{JFET,g}^{offset}$). Fig. S4 shows the response curves ($f_{out}$ vs. $f_{dr}$ for various $f_{mod}$) at different values of $V_{JFET,g}^{offset}$. It is observed that, as $V_{JFET,g}^{offset}$ decreases, the dynamic range of $f_{out}$ increases, indicating a reduction of $G_{shunt}$, consistent with the characteristics of the n-type JFET (2N4392, InterFET) used in this study.

When the data are replotted as a function of the input rate ratio $f_{dr}/f_{mod}$ (bottom panels), the degree of collapse of the input-output characteristics strongly depends on $V_{JFET,g}^{offset}$. An appropriate offset voltage yields a well-collapsed response onto a single curve, indicating recovery of the arithmetic invariant of the rate-coded DIVISION operator.

**S3. Calculation of the energy consumption for the image division operation**

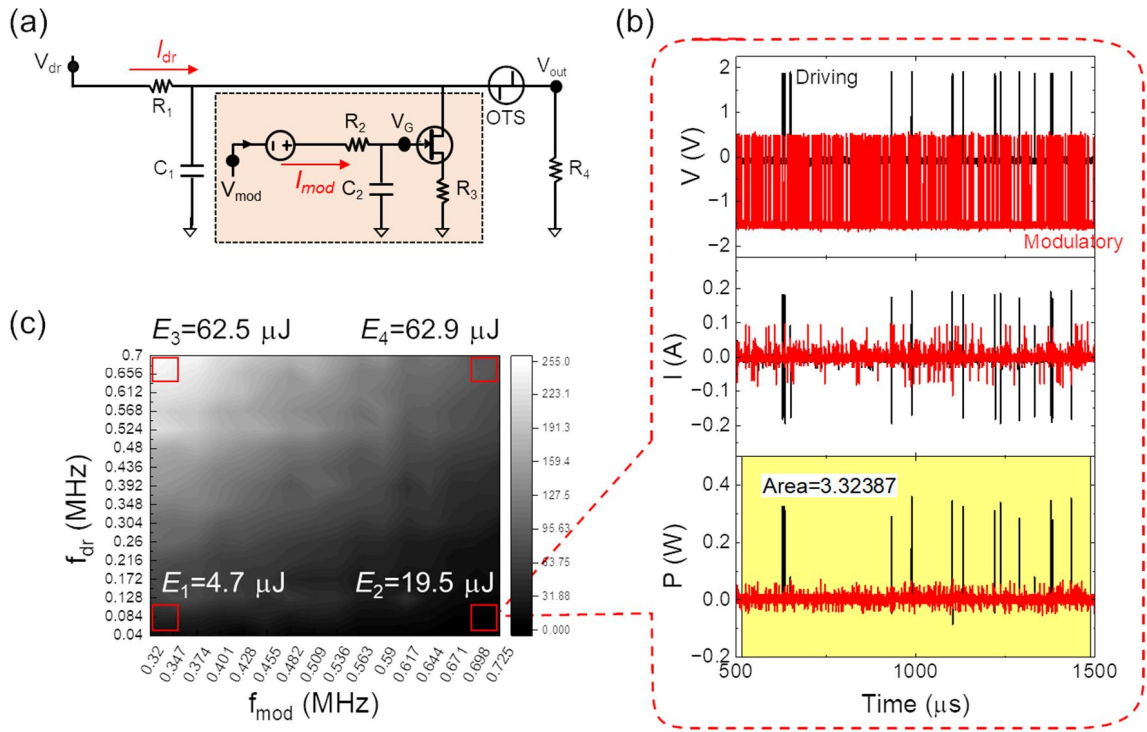

**Figure S3. Energy evaluation of the OTS-based DIVISION neuron under low-voltage operation.** (a) Schematic of the OTS-based DIVISION neuron used for energy measurements, indicating the driving and modulatory current paths ($I_{dr}$ and $I_{mod}$). (b) Representative Poisson train waveforms of the voltage ($V$, top panel), the current ($I$, middle panel), and the power ($P = V \times I$, bottom panel) of the driving (black) and the modulatory (red) sources during a selected time duration. The electrical energy consumed during a single rate-coded division operation is obtained by time-integrating the instantaneous power in both input paths over a fixed Poisson stimulation window. (c) To account for the dependence of energy consumption on input firing rates, four representative input combinations spanning the minimum and maximum values of the driving and modulatory rates ($f_{dr}$, $f_{mod}$) are selected ($E_1$ - $E_4$), as indicated on the ($f_{dr}$, $f_{mod}$) plane. The measured energies for these four cases are summarized, and their average value ($E_{avg} \approx 37.4$ µJ) is used as the representative energy per pixel-wise division under the low-energy operating condition ($V_{dr} = V_{mod} = 2$ V, $R_1 = 1$ kΩ).

## S4. Auger spectrum of a GSST film

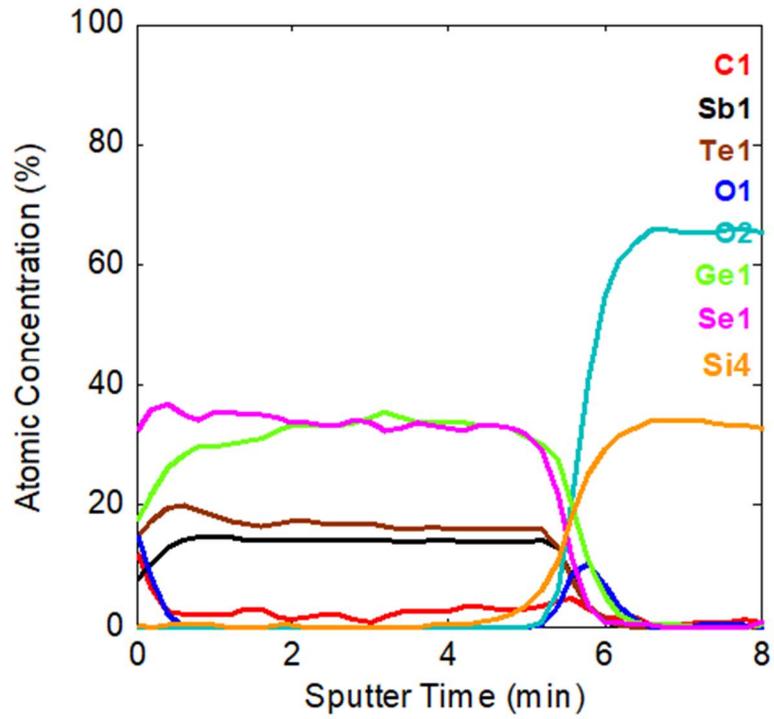

**Figure S4.** Depth profiles of atomic concentrations for Ge, Se, Sb, and Te in the $Ge_2Se_2Sb_1Te_{1.2}$ (GSST) switching layer measured by Auger electron spectroscopy (AES). The GSST film was deposited by RF co-sputtering using GeSe and $Sb_{2.5}Te_3$ targets.